\documentclass[aps,prd,preprintnumbers,nofootinbib,twocolumn]{revtex4}
\usepackage{bm}
\usepackage{latexsym}
\usepackage{dcolumn}
\usepackage{amsmath,amsfonts,amssymb}
\usepackage{graphicx,epsfig}
\usepackage{psfrag}
\usepackage{amsthm}

\def\be {\begin{equation}}
\def\ee {\end{equation}}
\def\bea {\begin{eqnarray}}
\def\eea {\end{eqnarray}}
\def\bc {\begin{center}}
\def\ec {\end{center}}
\def\bfg {\begin{figure}}
\def\efg {\end{figure}}
\def\bi {\begin{itemize}}
\def\ei {\end{itemize}}
\def\nn {\nonumber}

\def\la {\label}
\def\le {\left}
\def\ri {\right}

\def\no {\noindent}

\def\L  {\Lambda}

\def\beq{\begin{equation}}
\def\eeq{\end{equation}}
\def\br{\begin{eqnarray}}
\def\er{\end{eqnarray}}
\newcommand{\eel}[1] {\label{#1}\end{equation}}

\newcommand{\bdm}{\begin{displaymath}}
\newcommand{\edm}{\end{displaymath}}

\begin{document}
%\preprint{gr-qc/0704.xxxx}
%\hspace{15cm} 03/31/2009\\
\title{Quantum aether and an invariant Planck scale
%\footnote{Essay written for the Gravity Research Foundation 2010 Awards for Essays on Gravitation}
}

\author{Saurya Das $^1$} \email[email: ]{saurya.das@uleth.ca}
\author{Elias C. Vagenas $^2$} \email[email: ]{evagenas@academyofathens.gr}

\affiliation{$^1$~Theoretical Physics Group, Dept. of Physics and Astronomy,
University of Lethbridge, 4401 University Drive,
Lethbridge, Alberta, Canada T1K 3M4 \\}

\affiliation{$^2$~Research Center for Astronomy and Applied Mathematics,\\
Academy of Athens, \\
Soranou Efessiou 4,
GR-11527, Athens, Greece
}

\begin{abstract}
We argue that a quantum aether is consistent with the principle of relativity and
can provide an economical way of having an invariant quantum gravity or Planck scale.
We also show that it may change the effective scale at which quantum gravity effects
may be observable.
\end{abstract}
%\pacs{123xxx}

\maketitle

%%%%%%%%%%%%%%%%%%%%%%%%%%%%%%%%%%%%%%%%%%%%%%%%%%%%%%%

%$\ensuremath{^\circ}$

\null
%\vs{-0.46cm}
Predictions of certain theories of quantum gravity, such as the discreteness of space,
and requirements of certain others, such as the fundamental building blocks being strings,
are expected to be revealed
%existence of fundamental microscopic structures such as strings, are expected to show up
near the Planck scale (or the related string scale), characterized by the following length, mass and time
\footnote{These were first defined by Planck himself in \cite{planck}.}:
$
\ell_{Pl} = \sqrt{\frac{Gh}{c^3}} \approx 10^{-35}~m,~
m_{Pl} = \sqrt{\frac{h c}{G}} \approx 10^{-8}~kg,
%E_{Pl} &=& m_{Pl} c^2 \approx 10^{16} TeV,~
%T_{Pl} = \frac{E_{Pl}}{k} \approx 10^{32}~K,
%\ensuremath {^\circ}
t_{Pl} = \sqrt{\frac{Gh}{c^5}} \approx 10^{-44}~s$.
But since the above quantities are not Lorentz invariant, the scale of
fundamental structures or building blocks would appear to be different even in different
inertial frames in flat spacetimes and in vacuum.
Thus one is faced with the following dilemma, at least classically, that either \\
(i) Lorentz transformations are approximate and a suitable modification near
the Planck scale will make these frame independent, or \\
(ii)
%or there exists a preferred frame of reference, the rest frame of aether, in which
Planck quantities have the above values in a special frame. \\
The first possibility has been explored in the so-called Doubly Special Relativity (DSR) Theories,
where by realizing a particular non-linear version of the Lorentz group, it was shown that
both $c$ and $m_{Pl}$ can be made frame independent \cite{sm}
\footnote{In the rest of the paper, we will use $\hbar=1=c$ units.}.
However, incorporating the
new kinematics in a consistent dynamical theory has not been possible so far, and problems
of constructing macroscopic bodies from elementary particles satisfying energy momentum
conservation laws have proven difficult.
The second possibility on the other hand brings back the notion of luminiferous aether,
which as we know is fraught with problems, including contradictions with known experiments.

However, an early interesting proposal by Dirac in \cite{dirac1,dirac2} suggests that there
could yet be a third possibility, that of a quantum mechanical aether which does not violate
Lorentz symmetry. As we argue later, this in fact can also lend a precise meaning to an
invariant Planck scale.
Such an aether would presumably be light and made up of tiny constituents,
(as otherwise we would have detected its obvious presence) and hence subject to
the laws of quantum mechanics and the uncertainty principle. Thus it is natural to expect that the
ground state of such an aether (at any spacetime point) would be a uniform superposition of
all eigenstates of four-velocity
$v^\mu$ within the light cone at that point.
This is analogous (in three dimensions) to the %{\it spherically symmetric},
the $l=0$, or ground state of the hydrogen atom (without spin), which being a uniform superposition of all
possible position eigenstates, is itself a {\it spherically symmetric} state. Successive position
measurements on such identically prepared states would yield a probability distribution which is also
spherically symmetric. Similarly, the aforementioned state of aether, not associated with any specific
velocity, would be {\it Lorentz symmetric}!
More concretely, if $p^\mu$ is the four-momentum of the aether at any point
(which equal $m v^\mu$ if the constituents have a non-zero rest mass $m$),
then the wavefunction of aether at a point can be written as
%
%
%In this paper, following an earlier suggestion of Dirac \cite{dirac1,dirac2}, we
%show that there is a third possibility, that an aether treated quantum mechanically
%need not violate the Lorentz symmetry, and in fact can lend a precise meaning
%to the Planck scale.
%%
%Such an aether would presumably be very light, consisting of tiny
%constituents (as otherwise we would have detected its obvious
%presence), and thus be governed by quantum laws. Additionally,
%akin to the {\it spherically symmetric}, $l=0$, $s$-state of the
%hydrogen atom (without spin), Dirac assumed that the wavefunction of aether
%at any point in space and time would be a uniform superposition
%of all eigenstates of four-velocity $v^\mu$ within the light cone
%at that point.
%%\cite{dirac1,dirac2}.
%%
%%Without loss or generality, assuming that the particles of aether have a non-zero mass $m$,
%%and changing from the four velocities $v^\mu$
%%to
%Using the more commonly used four-momentum $p^\mu$ instead
%(which equals $m v^\mu$ if the constituents have a non-zero mass $m$)
%this state can be written
%%in the position basis
as
\bea
%\langle \vec r
| \Psi \rangle = N \int d\Omega_p~|\Psi_p\rangle \la{super}
\eea
where the integral is performed over a suitable Lorentz invariant measure $d\Omega_p$
over all momenta within the light cone, $|\Psi_p\rangle$ represents
a state of the aether with definite momentum $p^\mu$, and $N$ is the normalization constant.
This can be thought of as a four-dimensional generalization of the $s$-state, and
demonstrates (just like the hydrogen atom in a spherically symmetric state)
how in general quantum mechanics can produce an enhanced symmetry in certain states,
symmetry which is not present in the system's classical description.
This is a consequence of the superposition and uncertainty principles.
Note that as in the case of the Hydrogen atom,
and singlet states in various scenarios in particle physics,
the ground state of the aether is expected to be the perfectly
symmetric state (\ref{super}), and if perturbed by small amounts,
will return to this state rapidly.
Such small fluctuations can have observational
consequences however, and provide a mechanism of its detection
\footnote{
%\color{red}
The above notion of aether may in fact be compatible with, and
a concrete realization of
Einstein's efforts to re-introduce it, in which context he mentions
\cite{einstein}
{\it ``... the hypothesis
of ether in itself is not in conflict with the special theory of
relativity. Only we must be on our guard against ascribing a state
of motion to the ether.''}.
}.
%\vs{-.1cm}
Now the velocity of aether at any point is completely indeterminate {\it per se}, and
if a measurement of its velocity
is made, one of the infinite possibilities will be picked out at random
%\footnote{
%As pointed out in \cite{giulini}, decoherence may play an important role in this process.
%}.
This can be used to locally (in space and time)  define a rest-frame  in which the aether is at rest.
We go one step further and propose that if such a measurement is made at a spacetime point,  then
the aforementioned Planck scales are in {\it that frame}. This is similar in spirit to the
potential re-appearance of absolute time and absolute simultaneity in the presence of a quantum aether,
albeit in a small neighborhood and quasi-instantaneously, in accordance with the uncertainty principle,
and described statistically \cite{dirac2}.
Furthermore, just as observing an electron in a hydrogen atom initially in its $s$-state at a particular point in space
does not violate the spherical symmetry of the Hamiltonian, here too the random appearance of a preferred
frame does not violate the Lorentz symmetry observed in nature.
And as mentioned before, it would rapidly return to its quantum ground state.
And even if no such observation is made for a long time, one is still certain that
if and when it is made, it will result in a short-lived preferred frame.
Consequently, the existence of a fundamental scale (or more than one scale) is not incompatible with it.
Furthermore, the result would be the same when viewed from any inertial frame!
%Indeed Planck may have envisaged such an invariant notion of his constants.
Thus, as observed in \cite{dirac2}, it may constitute a {\it perfect vacuum}.
%In both the cases, the uncertainty principle guarantees that there is no contradiction with
%the underlying symmetry.
It is interesting to note that quantum mechanics
plays a crucial role in the definition of the Planck scale (which contains $h$) as well as in the above
proposed resolution.
We further note that the fundamental postulate of relativity, and in particular the
relativistic velocity addition formula ensures that
whatever random velocity of aether is measured at any spacetime point (say by a
mechanical experiment), this will not affect
the velocity of light in vacuum, and hence will have no effect on interference fringes
in Michelson-Morley type of experiments (which aimed to measure aether velocity)
\footnote{
Consider for example, an observer measures the aether velocity at any spacetime point to be
$\vec v$, and let the velocity of light measured by an observer who is at rest with respect to the aether
at that point be $\vec c$, with $|\vec c| = c= 2.9979 \times 10^8~m/s$. Then by the relativistic addition formula, the velocity of light as seen by the first observer is
$
\vec c~' = \frac{\vec v + \vec c \cdot \vec v~\hat v +
\sqrt{1-v^2/c^2}~(\vec c -\vec c \cdot \vec v~\hat v ) }{1 + \vec c~\cdot \vec v/c^2}~,
$
from which it is easily seen that $|\vec c~'|=c$.
}
. In other words, there
is no contradiction with known experiments.
%
%
%
%
%
%{
%\color{red}
Finally, we would like to point out that decoherence may play a very important role in the collapse of
wavefunction (\ref{super}), to a velocity (or momentum) eigenstate during a `measurement'.
This, as pointed out in \cite{giulini}, may be the outcome of an interaction of the aether with its environment,
and also of certain superselection rules, if the wavefunction has additional symmetries.
However, one would require more information, or need to make further assumptions about the aether and its
wavefunction to make a more detailed study. We leave this to a future publication.
%}

%
%Why would the aether be in the state described by (\ref{super})? Well, again as in the case of the Hydrogen atom,
%this perfectly symmetric state may well correspond to its ground, and hence most preferred state. Thus even if it is %perturbed by small amounts, it will have a tendency to return to this state rapidly. Such small fluctuations can have
%observational consequences though, and provide a mechanism of detecting the aether.

%Are the above
%We further note that since all possible aether velocities are equally likely to be observed, and the mean observed
%spatial velocity $\langle \vec v \rangle=0$, if the time scale of return of a measured aether
%state to its Lorentz symmetric normal state is considerably less than the typical time scales (light travel times) in
%Michelson-Morley type interference experiments (about 0.1 microsecond),  a rapid succession of velocity
%`measurements' (by the light beams) of neighboring aether particles (each in the superposed state (1))
%would yield a null result on the average. In other words, it would have no net effect on the interfering photons
%and their fringes and would not pick up a preferred frame, making it consistent with the observed results.
%More recent experiments may
%put tighter bounds on the aforementioned relaxation time, but a zero observed aether velocity would
%still be compatible with our proposed model.

%\vs{-.1cm}
What about the normalizability of the aether wavefunction?
As noted in \cite{dirac1} the wavefunction described by (\ref{super}) is ordinarily
not normalizable. It can be seen for example by substituting for $d\Omega_p$ a
relativistically invariant measure such as
$d^3p/(2\pi)^32E_p$ and evaluating
\bea
 \langle \Psi | \Psi \rangle =
\frac{|N|^2}{\pi} \int_0^\infty \frac{p^2 dp}{\sqrt{p^2 + m^2}} = |N|^2 \times \infty ~,
\la{integral}
\eea
which diverges for any non-zero $N$
(where we have used $\langle \Psi_{p'}|\Psi_p\rangle = 2E_p (2\pi)^3\delta(p'-p)$ \cite{peskin},
and $p=|\vec p|$.).
Dirac argued that since the state (\ref{super}) was after all an idealization, similar to plane waves
(which too are not normalizable), it
can be approached indefinitely close, but can never be attained in nature exactly.
However, as being argued here, if there is an invariant scale, one may also consider it as
an upper cut-off $\Lambda$ (with or without an invariant lower cut-off), and then
for the choice $N = \sqrt{a\pi}/\L$, and $N = (3m\pi /\Lambda^3)^{1/2}$ for $\Lambda \gtrsim m$
and $\Lambda \ll m$ respectively (again, $m$ being the mass of the aether quanta, or that of its
fundamental constituents, if thought of as a fluid)
the integral in (\ref{integral}), which go as $\L^2$ and $\Lambda^3/m$ in these limits, is finite
\footnote{$a=2$ for $\Lambda \gg m$ and
$a=\le[ 1/\sqrt{2}-1/2\ln(1+\sqrt{2}) \ri]^{-1}\approx 3.7$ for $\Lambda \approx m$.}.
And with a normalizable wavefunction, the aether state of perfect vacuum could well be
attainable in nature.
Note that within the current interpretation, $\Lambda$ being Lorentz invariant,
the integral can be regarded as Lorentz invariant as well.
Note also that this cut-off could be the Planck scale.
But another intermediate scale can exist by the same token \cite{advplb}.
In any case, such a fundamental scale is expected to arise from a correct theory of quantum gravity
or of yet unknown physics beyond the electroweak energy scale.
This does not seem to be the case for DSR theories.

Finally, we ask the question: since the aether's
motion is random at any point, what would be the effective Planck
scale at which quantum gravity effects would be expected? First we
note, following Lorentz length contraction, time dilation etc,
that the numbers presented at the beginning provide strict lower
bounds (for Planck mass and Planck time) and an upper bound (for
Planck length). One can then compute the average of the `observed'
Planck mass, namely  $M_{Pl} = \gamma m_{Pl}$ with
$\gamma=(1-v^2/c^2)^{-1/2}=\sqrt{p^2+m^2}/m$, as
\bea
\langle M_{Pl} \rangle &=&
 \langle \Psi | ( m_{Pl}\gamma ) |\Psi \rangle =
\frac{|N|^2}{\pi} m_{Pl} \int_0^\L \frac{p^2 \gamma dp}{\sqrt{p^2 + m^2}}~~~
%= |N|^2 \times \infty ~,
~\la{integral2} \\
&=& \frac{a~m_{Pl}}{3} \frac{\Lambda}{m} ~,~ \Lambda \gtrsim m
\nn \\
&=&
{m_{Pl}} ~,~\Lambda \ll m
\nn
\eea
which is finite. For the first case, $\Lambda \gtrsim m$, we have at least two possibilities \\
(i) $\Lambda \gg m $, i.e. an aether which is very light compared to the cutoff scale, the latter being
for example, the Planck scale. Then $\langle M_{Pl} \rangle$ would be huge, making it
(as well as quantum gravity effects) impossible to observe, and practically
irrelevant.
 \\
(ii) $\Lambda \approx m$, with $\Lambda$ being much less than the Planck scale, as otherwise
the aether would be heavy. Then $\langle M_{Pl} \rangle \approx m_{Pl}$ and the scale of
quantum gravity effects remain unchanged. \\
For the second case, $\Lambda \ll m$, too, the quantum gravity scale remains unchanged.
%\footnote{
%The third possibility $m >> \Lambda$, with enormously lowered quantum gravity scales is ruled out
%because of the lower bound on observed Planck mass mentioned before. This would also require an
%unnaturally small cutoff $\Lambda$.
%}
%\\
%(iii) $m >> \Lambda$, and quantum gravity phenomena may be observable at much lower scales which is
%impossible since $m_{Pl}$ sets a strict lower bound. \\
%%
Although possibility (i) above may appear more natural, in
the end, it is for experiments to decide whether one of the above is true.
Experiments may also help to obtain bounds on the cut-off or the aether mass, using
(\ref{integral2}).

%\noindent
%(ii) that the constituents of aether are massless,
%%$v=c$ in (\ref{super}), i.e.
%the measured velocity of aether will always be the speed of light,
%pointing randomly in any direction along the boundary of the light cone at any point.
%Furthermore, if they are all assumed to be (approximately) of the
%Then the
%integration in (\ref{super}) just gives $4\pi$, and further assuming $\langle \Psi_p | \Psi_{p'}\rangle=1$, we get
%$N=1/\sqrt{4\pi}$. However, in this case, it is {\it a priori} not clear what a finite value of Planck length or mass
%in the luminal frame will translate to in a sub-luminal one.

To conclude, we have shown here that a quantum mechanical version of an all permeating aether
is consistent with the principle of relativity, % (there is no need to change it!),
and may provide a mechanism for having an invariant
quantum gravitational scale.
%,  and as a bonus, invariant cut-off scales in quantum field theories as well.
Further, depending on the relative magnitudes of the mass of its quanta and this cut-off,
the effective Planck scale may acquire very large or small values.
There is no need to change the theory of relativity.
The simplicity of its construction ensures that problems associated with the
composition of macroscopic bodies or with
% satisfying relativity principles,
additivities of energy and momentum etc do not arise. Neither does it
present any obvious problem to existing dynamical theories such as gauge theories and
general relativity. Thus it may represent the physical vacuum
\footnote{
%{\color{red}
Again, according to Einstein \cite{einstein}
{\it ``To deny the ether is ultimately to assume that empty space has no
physical qualities whatever. The fundamental facts of mechanics
do not harmonize with this view.''} and
{\it
``According to the general theory of
relativity space without ether is unthinkable;...''}.
%}
}.
A consistent dynamical theory of aether itself
would have to be formulated however, which would shed light on the nature of its constituents, and the fate of observable
preferred frames in the long run.
It is tempting to speculate that this version of aether may have some bearing on the abundance of
{\it Dark Matter} and {\it Dark Energy}
in our universe. One would of course have to compare with their quantitative estimates and also
reconcile with the fact that Dark Energy is associated with a minuscule vacuum energy density.
%or cosmological constant.
Such a cosmological connection with an aether which is also treated as a
superfluid was proposed in \cite{sudarshan}. Aether in the cosmological context has also
been recently studied in \cite{newaetherpaper}.
Naturally, one must also check the consistency of our current hypothesis with all other relevant
physical theories, and more importantly, extract predictions which can potentially be checked in the laboratory or
in astrophysical observations. This would be a true test of its existence.

%\vs{.5cm}
%%%%%%%%%%%%%%%%%%%%%%%%%%%%%%%%%%%%%%%%%%%%%%%%%%%%%%
\no {\bf Acknowledgment}

SD thanks R. Sorkin for discussions. We thank the anonymous
referees for their useful comments which helped improve the paper.
This work was supported in part by the Natural Sciences and
Engineering Research Council of Canada and by the Perimeter
Institute for Theoretical Physics.
%%%%%%%%%%%%%%%%%%%%%%%%%%%%%%%%%%%%%%%%%%%%%%%%%%%%%
\\
%\par\noindent
%\line(1,0){40}
%{\footnotesize
%\par\noindent
%$^{6}$
%{\color{red}
%Again, according to Einstein \cite{einstein}
%{\it ``To deny the ether is ultimately to assume that empty space has no
%physical qualities whatever. The fundamental facts of mechanics
%do not harmonize with this view.''} and
%{\it
%``According to the general theory of
%relativity space without ether is unthinkable;...''}.}
%}
%%%%%%%%%%%%%%%%%%%%%%%%%%%%%%%%%%%%%%%%%%%%%%%%%%%%%%%%%

%$$
%\vec c~' = \frac{\vec v + \vec c \cdot \vec v~\hat v +
%\sqrt{1-v^2/c^2}~(\vec c -\vec c \cdot \vec v~\hat v ) }{1 + \vec c~\cdot \vec v/c^2}~,
%$$

%\section*{References}

\end{document}